\documentclass[aps,prl,twocolumn,superscriptaddress,notitlepage,nofootinbib,longbibliography]{revtex4-2}
 \usepackage{mathrsfs}
 \usepackage{epsfig}
 \usepackage{graphicx}
 \usepackage{amsfonts}
 \usepackage[figuresright]{rotating}
 \usepackage{amssymb}
 \usepackage{amsmath}
 \usepackage{dcolumn}
 \usepackage{bm}
 \usepackage{xcolor}
 \usepackage{color}
 \usepackage{braket}
 \usepackage{units}
 \usepackage{xspace}

 \usepackage{bbold}
 \definecolor{mypink3}{cmyk}{0, 0.7808, 0.4429, 0.1412}
 \definecolor{mypink1}{rgb}{0.858, 0.188, 0.478}
 \definecolor{mypink2}{RGB}{219, 48, 122}
 \usepackage[colorlinks=true, allcolors=mypink1]{hyperref}
 \usepackage[fleqn]{mathtools}

 \usepackage[shortlabels]{enumitem}
 \newcommand{\ECNU}{Quantum Institute for Light and Atoms, Department of Physics, School of Physics, East China Normal University, Shanghai 200062, China}
 \newcommand{\SBH}{Shanghai Branch, Hefei National Laboratory, Shanghai 201315, China}
 \newcommand{\SPA}{School of Physics and Astronomy,  Shanghai Jiao Tong University, Shanghai 200240, China}
  \newcommand{\TLI}{Tsung-Dao Lee Institute, Shanghai Jiao Tong University, Shanghai 200240, China}
 \newcommand{\SRC}{Shanghai Research Center for Quantum Sciences, Shanghai 201315, China}
 \newcommand{\CIC}{Collaborative Innovation Center of Extreme Optics, Shanxi University, Taiyuan, Shanxi 030006, China}

\begin{document}

\title{Atomic Regional Superfluids in Two-Dimensional Moiré Time Crystals}
	
 \author{Weijie Liang}
 \affiliation{\ECNU}
 \author{Weiping Zhang}
 \email{wpz@sjtu.edu.cn}
 \affiliation{\SPA}
  \affiliation{\TLI}
 \affiliation{\SBH}
 \affiliation{\SRC}
 \affiliation{\CIC}
 \author{Keye Zhang}
 \email{kyzhang@phy.ecnu.edu.cn}
 \affiliation{\ECNU}
 \affiliation{\SBH}

\begin{abstract}

Moiré physics has transcended spatial dimensions, extending into synthetic domains and enabling novel quantum phenomena. We propose a theoretical model for a two-dimensional (2D) moiré time crystal formed by ultracold atoms, induced by periodic perturbations applied to a nonlattice trap. Our analysis reveals the emergence of regional superfluid states exhibiting moiré-scale quantum coherence across temporal, spatial, and spatiotemporal domains. This work provides fundamental insights into temporal moiré phenomena and presents an alternative pathway to engineer spatial moiré phases without requiring twisted multilayer lattices.

\end{abstract}
\maketitle

\emph{Introduction}---Twistronics has revolutionized quantum materials science through interlayer rotation in 2D systems, enabling discoveries from unconventional superconductivity in twisted bilayer graphene \cite{cao2018unconventional,yankowitz2019tuning,kerelsky2019maximized,lu2019superconductors,choi2019electronic,liu2021tuning} to topologically protected quantum devices \cite{spanton2018observation,xie2021fractional,kezilebieke2022moire,cai2023signatures} and novel quantum phases in moiré lattices \cite{tong2017topological,cao2018correlated,sharpe2019emergent,codecido2019correlated,chen2019evidence,tran2019evidence,zheng2020unconventional,zheng2024superconductivity}. While its quantum simulations using atomic Bose-Einstein condensates (BECs) in moiré optical lattices \cite{meng2023atomic} and photonic moiré crystals \cite{wang2020localization,fu2020optical,lou2022tunable,luan2023reconfigurable} have surpassed solid-state parameter control limits, these approaches remain constrained by two fundamental requirements, physical twisted lattice structures and exclusive focus on spatial dimensions. Recent demonstrations of frequency-domain moiré lattices \cite{yu2023moire} suggest twistronics' potential in synthetic dimensions. However, the temporal moiré lattices and the intrinsic link between spatial and temporal moiré physics in quantum many-body systems remain unexplored. Specifically, temporal moiré phases are expected to offer advantages to quantum applications, analogous to those provided by their spatial counterparts but operating in the time domain \cite{montenegro2023quantum,cabot2024continuous,iemini2024floquet,chen2025phase,yousefjani2025discrete,gribben2025boundary,dumitrescu2022dynamical}.

We present a lattice-free scheme for unified spatial and temporal twistronics simulations using a moving BEC within a deeply confined 2D potential well subjected to multifrequency “bobbing” modulation. This approach eliminates the requirement for complex multilayer lattice potentials \cite{meng2023atomic,gonzalez2019cold} or engineered interlayer tunneling \cite{salamon2020simulating}, while generating both spatial and temporal moiré quantum phases. These phases exhibit characteristic regional superfluidity with moiré-patterned coherence distributions, where crucially, the effective twist angles, layer numbers, and resulting spatiotemporal moiré patterns can be dynamically tuned through precise frequency control, offering unprecedented flexibility in quantum simulation.

Building on this platform, we develop a 2D moiré time crystal framework within Floquet phase space (FPS) that offers a complete understanding of these quantum phenomena, surpassing conventional approaches like period-doubling oscillation \cite{PhysRevLett.47.1349} or dynamical localization \cite{PhysRevA.79.013611} which only partially capture temporal or spatial behaviors. In this space, where time effectively emulates spatial dimensions to host exotic quantum phases previously confined to spatial systems \cite{sacha2015anderson,autti2018observation,giergiel2018time,giergiel2019topological,kopaei2022topological,guo2022phase,kopaei2024topologically}, the moiré lattice enables seamless interconversion between spatial and temporal roles while maintaining essential moiré characteristics. This approach not only circumvents the need for physical twisted lattices but also uncovers novel spatiotemporal moiré phases, establishing time as an engineering degree of freedom for advanced twistronics applications.

\begin{figure}[t]
	\centering
	\includegraphics[width=0.95\linewidth, height=0.6\linewidth]{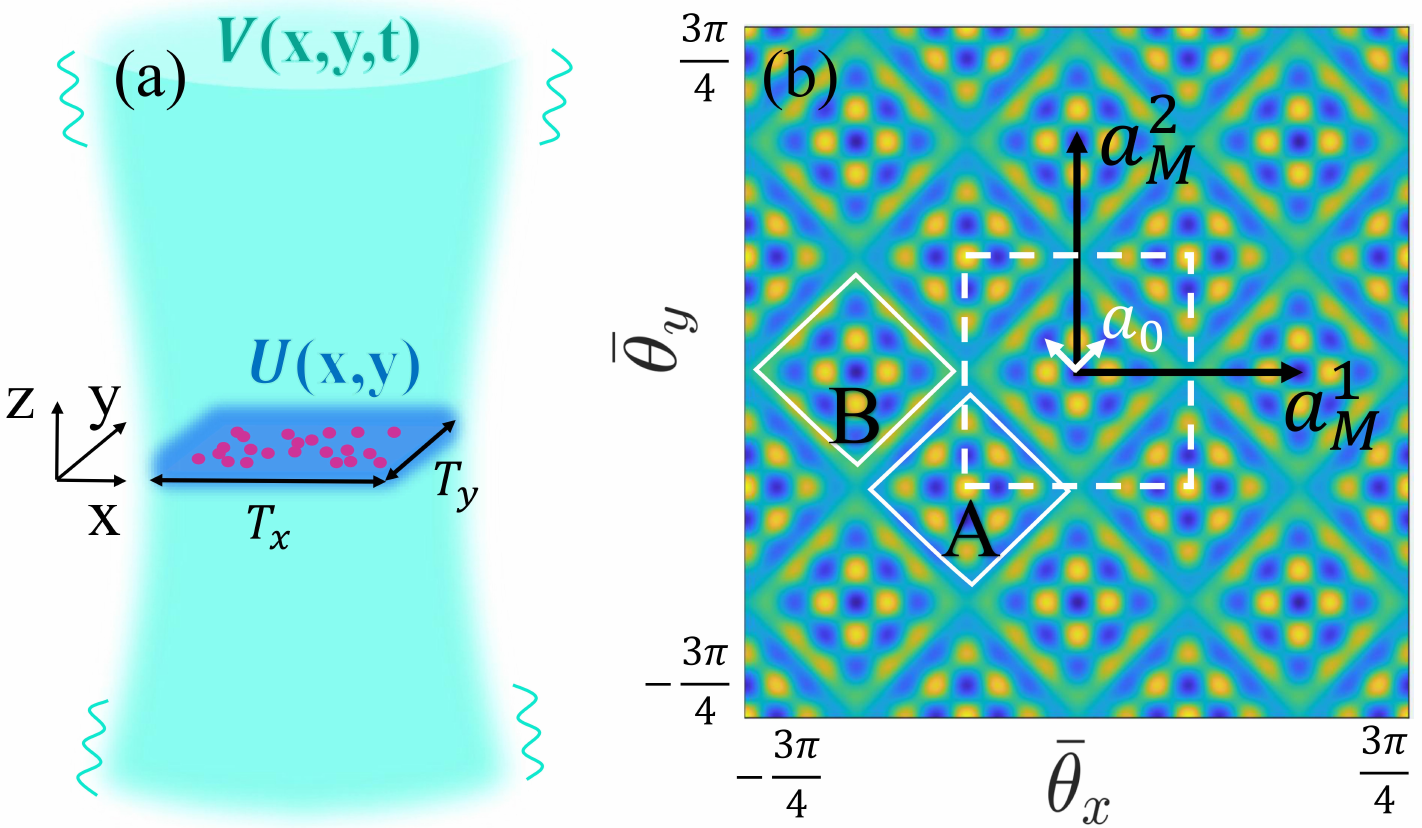}
	\caption{(a) A schematic diagram of the model depicts ultracold atoms moving within a 2D square potential well (blue) and experiencing perturbation from an additional laser beam (light blue) featuring multifrequency intensity modulation. (b) 2D FPS effective potential $\mathcal{V}_{n=16,m=12}(\bar\theta_x, \bar\theta_y)$, with the depth indicated by color, at twist angle $\alpha\approx16.26^\circ$ and primary cell periodicity $a_0=\pi/10$ denoted by the white shorter arrows. Two distinct morié patterns, $A$ and $B$, are observed (bordered by white dashed lines), corresponding to a quadrilateral unit cell with equal-length morié vectors $|\mathbf{a_{M}^{1}}|=|\mathbf{a_{M}^{2}}|=5a_0$.}
	\label{fig1}
\end{figure}

\emph{Model}---We consider a system of ultracold atoms confined in a two-dimensional deep potential well $U(x,y)$, as depicted in Fig.~\ref{fig1}(a). Rather than directly constructing additional twisted-bilayer lattice potentials using optical lattices of different orientations, we subject the atoms to a weak multi-frequency perturbation $V(x,y,t)$. The single-atom Hamiltonian in 2D is
\begin{equation}
    \label{Eq1}
    H_{sa} = \frac{p_{x}^2 + p_{y}^2}{2m_{a}} + U(x,y) + V(x,y,t) 
\end{equation}
where $p_{x}$ and $p_{y}$ are the atomic momenta along the $x$ and $y$ directions, respectively, and $m_{a}$ is the atomic mass.
When a frequency component of $V$ matches an integer multiple of the atomic orbital frequency in $U$, a lattice structure emerges in the FPS, as predicted by canonical perturbation theory. The resulting periodic quantum dynamics are captured by a quasienergy band approach, akin to recent studies on time crystals \cite{guo2013phase,sacha2015modeling,matus2019fractional}. By introducing multiple perturbation frequencies and carefully establishing resonant relations, we can create twisted bilayer lattices or even more complex lattice structures within the FPS.

To be concrete, we take $U$ as a 2D infinite square well of width $L$, and $V(x,y,t) = V_0 x^2 y^2 \sum_l f_l \cos\omega_l t$ as a quadratic potential modulated at frequencies $\omega_l$. The unperturbed atomic motion is periodic along $x$ and $y$, with the orbital frequencies $\Omega_{x,y} = 2\pi/T_{x,y} = (\pi/L) \sqrt{2E_{x,y}/m_a}$, where $E_{x,y}$ is the initial kinetic energy. In action-angle coordinates $(I_{x,y}, \theta_{x,y})$ defined by $\theta_x = \text{sign}(p_x)(\pi x/L + \pi/2)$ and $I_x = L\sqrt{2m_a E_x}/\pi$ (similarly for $y$) \cite{PhysRevA.107.022605}, the perturbation potential admits a Fourier expansion,
\begin{equation}
V(x,y,t) = \sum_{\mathfrak{n},\mathfrak{m},l} \mathcal{V}_{\mathfrak{n}\mathfrak{m}l} e^{i (\mathfrak{n}\theta_x + \mathfrak{m}\theta_y \pm\omega_l t)}.
\end{equation}

The secular approximation dictates that only resonant terms, satisfying $\mathfrak{n}\theta_x + \mathfrak{m}\theta_y\pm \omega_l t = 0 $ ($  \mathfrak n,  \mathfrak m \in \mathbb{Z}$), significantly influence the long-term dynamics, while nonresonant terms average out due to rapid oscillation. Considering the angle evolution $\theta_{x,y} \approx \Omega_{x,y} t$ for a weak perturbation, we apply four perturbation frequencies tuned to $\omega_1=n\Omega_x + m\Omega_y$, $\omega_2 = m\Omega_x + n\Omega_y$, $\omega_3 = n\Omega_x - m\Omega_y$, and $\omega_4 = m\Omega_x - n\Omega_y$ (with $n, m \in \mathbb{Z}$). By transforming to the comoving frame via $\bar{\theta}_{x,y} = \theta_{x,y} - \Omega_{x,y} t$, we obtain a static lattice Hamiltonian in the FPS, as shown in Supplemental Material 
\cite{supplementarymaterial}\nocite{buchleitner2002non,lichtenberg2013regular,sacha2016anderson,sacha2015modeling,peng2018time,martin2017topological,casati1989anderson,ho1983semiclassical,ikeda2021fermi,zhao2021random,lee2002energy},
\begin{equation}
    \label{Eq2}
    \begin{aligned}
    \begin{split}   \mathcal{H}_{sa}&=\frac{P^{2}_{\bar{\theta}_{x}}+P^{2}_{\bar{\theta}_{y}}}{2m_{\rm eff}}+\mathcal{V}_{nm}[\cos(m\bar{\theta}_{x}+n\bar{\theta}_{y})+\cos(n\bar{\theta}_{x}-m\bar{\theta}_{y})\\
    &+\cos(n\bar{\theta}_{x}+m\bar{\theta}_{y})
    +\cos(m\bar{\theta}_{x}-n\bar{\theta}_{y})]\,,
    \end{split}
    \end{aligned}
\end{equation}
where $P_{\bar{\theta}_{x,y}} = I_{x,y}^\prime - I_{x,y}^0$ represents the deviation from the unperturbed action, serving as the canonical momentum conjugate to $\bar{\theta}_{x,y}$ and $m_{\mathrm{eff}}=\frac{m_a L^2}{\pi^2}$ is the effective mass. 
The superposition of these four equally weighted cosine terms directly constructs an effective twisted bilayer square lattice.

The resulting moiré superlattice, illustrated in Fig. \ref{fig1}(b), features two distinct square patterns (A and B) with the central region of A forming a barrier and B a deep well. Its geometry is characterized by a primary period $a_0 = 2\pi/\sqrt{n^2 + m^2}$, a twist angle $\alpha = \left| \arctan\left( \frac{m}{2n} - \frac{n}{2m} \right) \right|$, and a moiré period $|\mathbf{a}_{M}^{1,2}| = \bar{a} / [2 \sin(\alpha/2)]$, where $\bar{a}=\sqrt{2}a_0$ if $(m+n)/s$ is odd and $\bar{a}=a_0$ if even, with $s$ being the greatest common divisor of $m$ and $n$. The commensurability condition required for the formation of moiré patterns is ensured when $(n, m)$ form a Pythagorean triple. Remarkably, the absence of a physical lattice potential enables dynamic tuning of the moiré geometry, including orientation, twist angle, and layer number, by simply adjusting the frequency distribution of $V$, facilitating the realization of diverse and complex configurations \cite{supplementarymaterial}.

The validity of this lattice Hamiltonian requires suppressing contributions from accidental and near-resonant terms. This is achieved by selecting incommensurate base frequencies $\Omega_x / \Omega_y$ and a perturbation profile that inherently attenuates higher-order couplings \cite{PhysRevLett.119.230404}.
The quartic potential $V \propto x^2 y^2$ used here exemplifies the latter, as its Fourier coefficients $\mathcal{V}_{\mathfrak{n}\mathfrak{m}l} = 4V_0 L^4 f_l / (\pi^4 \mathfrak{n}^2 \mathfrak{m}^2)$ decay rapidly with increasing $|\mathfrak{n}|, |\mathfrak{m}|$. Furthermore, the perturbative treatment requires the dimensionless strength $2V_0 L^2 / (\pi^2 m_a \Omega_x \Omega_y)$ to remain small, as detailed in \cite{supplementarymaterial}. Any alternative potential must retain an $x$-$y$ coupled structure to maintain interdependence between the motions along the two directions, which is essential for controlling the lattice orientation in the FPS.

Unlike moiré lattices in conventional 2D phase space formed by noncommutative coordinates $x$ and $p_x$ \cite{guo2024engineering}, our lattices are spanned by commutative coordinates $\bar{\theta}_x$ and $\bar{\theta}_y$. This key distinction allows the realization of rich quantum phases, similar to those found in 2D spatial moiré systems, especially when atom-atom interactions are included.
Furthermore, the analogous roles of $x$ and $t$ in the coordinate transformation $\bar\theta_x = \text{sign}(p_x)(\frac{\pi x}{L} + \frac{\pi}{2}) - \Omega_x t$ (and similarly for $\bar\theta_y$) highlight the dual manifestation of these FPS moiré phases.
These phases can appear not only in the spatial domain at a fixed time, even in the absence of a spatially twisted lattice potential, but also in the temporal domain at a fixed position, revealing distinct temporal correlations in the dynamics.

Specifically, at a fixed time $t = t_0$, a moiré lattice potential resembling Fig.~\ref{fig1}(b) appears in real space $(x,y)$, with a primary period $a_0^r = a_0 L/\pi$ and a moiré period $a_M^r = a_M L/\pi$. 
Conversely, at a fixed spatial position $(x_0, y_0)$, the temporal dynamics can exhibit effective two-dimensional behavior when the orbital frequencies are widely separated (e.g., $\Omega_x \gg \Omega_y$ or vice versa). This allows time to be discretized using two distinct measurement intervals $\Delta t_x \ll \Delta t_y$ as 
\begin{equation}
   t=\Delta t_x j +\Delta t_y k ,
   \label{tjk}
\end{equation} 
where $j,k$ are positive integers. 
Substituting this into expressions for $\bar\theta_{x,y}$ and choosing $\Delta t_y = T_x$ (i.e., $\Delta t_y \Omega_x = 2\pi$), we obtain
\begin{eqnarray}
    \bar{\theta}_x &=& \theta_{x_0} - \Omega_x \Delta t_x j - 2k\pi, \label{thetaj}\\
    \bar{\theta}_y &=& \theta_{y_0} - \Omega_y \Delta t_y k - \delta_j,
    \label{thetak}
\end{eqnarray}
where $\bar{\theta}_x$ varies linearly with $\Delta t_x j$ and remains independent of $\Delta t_y k$, while $\bar{\theta}_y$ varies linearly with $\Delta t_y k$ with a negligible offset $\delta_j = \Omega_y \Delta t_x j$ under the condition $\Omega_x \gg \Omega_y$.
The 2D time degrees of freedom are represented by discrete variables $(j,k)$, but here $\Omega_x \Delta t_x, \Omega_y \Delta t_y \ll 2\pi$ ensure sufficient resolution for capturing the moiré lattice structure in the time domain, with primary periods $T_0^{x,y}=a_0/\Omega_{x,y}$ and moir\'e periods $T_M^{x,y}=a_M/\Omega_{x,y}$.

Furthermore, based on a similar principle, an even more intriguing spatiotemporal moiré lattice can arise when we fix just one spatial coordinate and utilize a specific time interval. For example, by fixing $y = y_0$ and $t = \Delta t_y k$, we obtain
\begin{eqnarray}
    \bar{\theta}_x &=& \theta_x - 2\pi k, \\
    \bar{\theta}_y &=& \theta_{y_0} - 2\pi \left(\frac{\Omega_y}{\Omega_x}\right) k,
\end{eqnarray}
which results in an effective two-dimensional spatiotemporal coordinate system $(x, k)$, from which the spatiotemporal moiré lattices and moiré phases emerge. Similarly, an effective $(j,y)$ system can be derived.

\emph{Regional superfluid}---Under repulsive atom-atom interactions, the cold atoms can form a characteristic quantum moiré phase, regional superfluid in the FPS. This phase exhibits moiré density patterns where coherence is confined within each pattern and breaks down at the boundaries due to sharp potential changes.

The spacetime symmetry of the FPS coordinates ($\bar{\theta}_{x},\bar{\theta}_{y}$) allows this unique moiré phase to manifest in spatial, temporal, or spatiotemporal domains. The phase can be studied through measurements taken at strategically chosen instants or positions.
To validate this picture, we simulate the atomic dynamics using the Gross-Pitaevskii equation in the laboratory frame,
\begin{equation}
\label{Eq6}
i\hbar\frac{\partial\psi}{\partial t} = \left(H_{sa}(t) + g\eta N|\psi|^{2}\right)\psi,
\end{equation}
where $\psi$ denotes the condensate wave function (initially a wave packet with given momentum), $N$ the total number of atoms, and $g=4\pi\hbar^2a_s/m_a$ the 3D atomic interaction strength.
The coefficient $\eta$ quantifies the dimensional reduction of interactions from 3D to 2D.
It takes a constant value, $\eta^{-1} = \sqrt{h / ( m_a \omega_z)}$, when the axial confinement frequency $\omega_z$ satisfies $\Omega_{x,y} \ll \omega_z \ll \hbar / (m_a a_s^2)$. This condition ensures that the $z$-axial motion is frozen into the ground state while remaining in the weakly interacting regime. Beyond this parameter range, an energy-dependent $\eta$ must be used, as detailed in \cite{lee2002energy, supplementarymaterial}.
The mean field nonlinearity induced by atomic interaction is crucial for both producing exotic quantum many-body phases and breaking $H_{sa}(t)$'s discrete time translation symmetry, essential for realizing time crystals.

\begin{figure}[t]
	\centering
	\includegraphics[width=0.95\linewidth, height=0.6\linewidth]{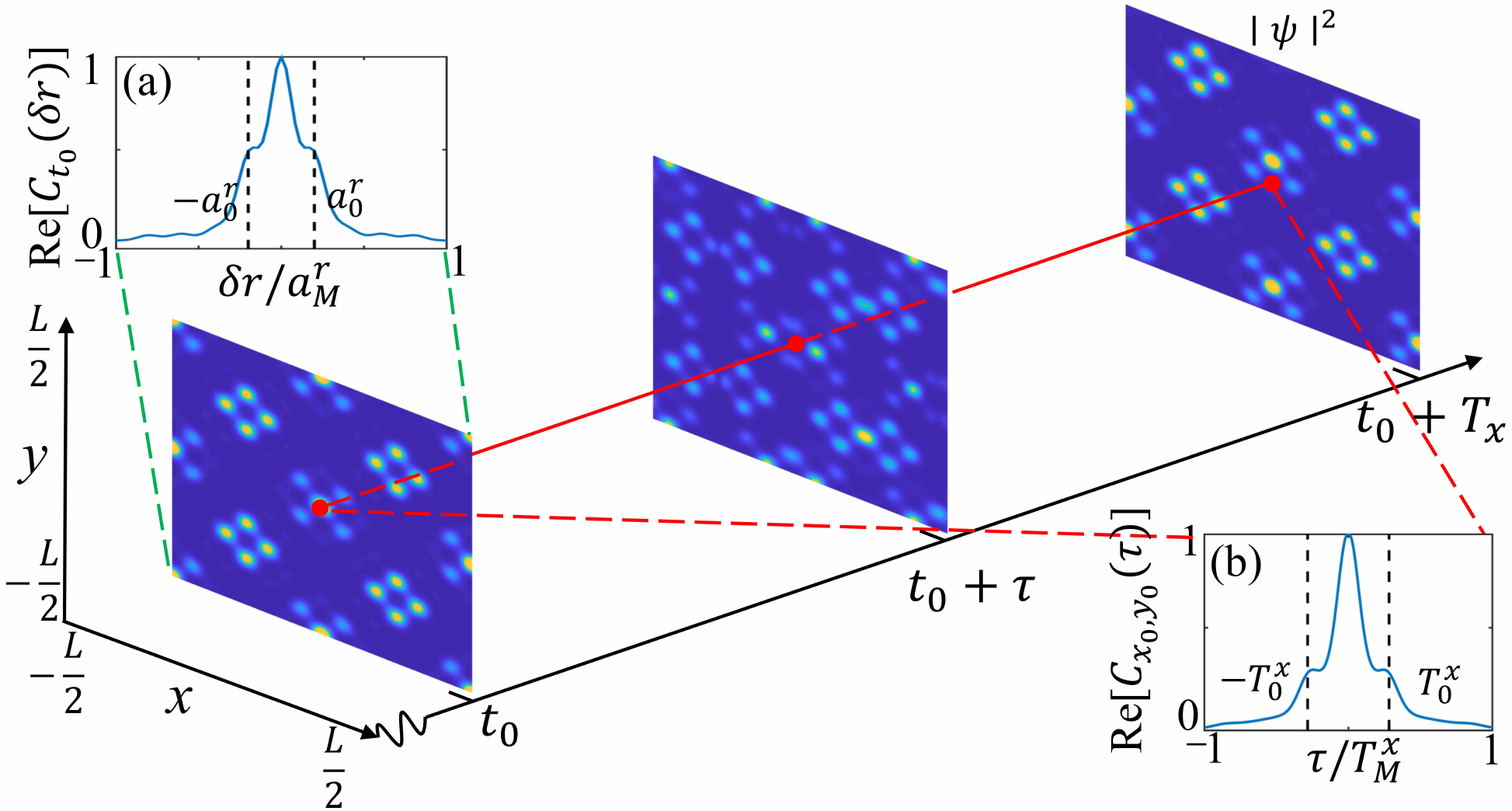}
    \caption{The atomic probability density $|\psi(x,y,t)|^2$ at three moments is accompanied by the inset showing (a) the real part of spatial autocorrelation $C_{t_0}(\delta r)$ at $t_0=98T_y$ within a spatial moir\'e period $a_M^r$ (dashed lines mark spatial primary cell period $a_0^r$) and (b) the real part of temporal autocorrelation $C_{x_0,y_0}(\tau)$ at position $(x_0,y_0)$ (red dot) within a temporal moir\'e period $T_M^x=T_x/4$ (dashed lines mark the temporal primary cell period $T_0^x=T_x/20$), using parameters $\Omega_{x}=\pi\times10^{3}E_{r}/\hbar$, $\Omega_x/\Omega_{y}=64\pi$, $\mathcal{V}_{16,12}=0.35E_{r}$, $E_{r}=\frac{\hbar^{2}}{m_{a}}(\frac{2\pi}{a_0^r})^2$, and $\eta gN=10^2\hbar^{2}/m_{a}$.}
	\label{fig2}
\end{figure}

We first confirm the spatial regional superfluid. At times $t_0$ that are integer multiples of $T_x$ or $T_y$, the atomic density and coherence in laboratory coordinates $(x, y, t_0)$ closely match those in FPS, up to minor displacements from the coordinate transformation. At these instants, the system exhibits clear regional superfluidity with moir\'e-patterned density distributions in real space (as the slices in Fig.~\ref{fig2}), despite the absence of a physical moiré lattice potential.

We quantify this using the spatial autocorrelation function,
\begin{equation}
    C_{t_0}(\delta \boldsymbol{r}) = \int \psi^*(\boldsymbol{r},t_0)\psi(\boldsymbol{r}+\delta\boldsymbol{r}, t_0) d^2\boldsymbol{r}
    \label{Cr}
\end{equation}
with time $t_0$ fixed and $\delta\boldsymbol{r}$ denoting 2D spatial separation distance vector.
For simplicity, we consider the separation along $\delta y = \delta x$, and the results are presented in the inset of Fig.~\ref{fig2} (a). 
With increasing separation, $C_{t_0}(\delta \boldsymbol{r})$ exhibits a crossover in its decay behavior, superimposed on density-dependent oscillations. The decay profile is approximated within the Bogoliubov and moiré band theories by $\sim e^{-k_\Delta^r |\delta \boldsymbol{r}|} / |\delta \boldsymbol{r}|$, yielding a coherence length $1/k_\Delta^r \approx a_0^r \sqrt{2J/\Delta_M}$ in the weak-interaction limit \cite{supplementarymaterial}, where $\Delta_M$ is the lowest moiré band gap and $J$ is the intercell tunneling rate. For our parameters, $1/k_\Delta^r \approx 2.5a_0^r \approx 0.5a_M^r$. At separations shorter than this length, power-law decay prevails, characteristic of superfluid, while beyond this length exponential decay dominates, confirming a loss of long-range coherence and the regional nature on the scale of the moiré period $a^r_{M}$.

Analogously, we validate the temporal regional superfluid via the temporal autocorrelation function at a fixed position $(x_0,y_0)$,
\begin{equation}
C_{x_0,y_0}(\tau)=\int \psi^*(x_0,y_0,t)\psi(x_0,y_0,t+\tau)dt.
\label{Ct}
\end{equation}
Figure~\ref{fig2}(b) shows prominent coherence within about two primary period $T_0^x$, which vanishes for $\tau > T_M^x$, mirroring the spatial behavior. The longer-time structure with periods $T_0^y$ and $T_M^y$ is not visible on this timescale.

\begin{figure}[t]
	\centering
\includegraphics[width=0.95\linewidth,height=0.6\linewidth]{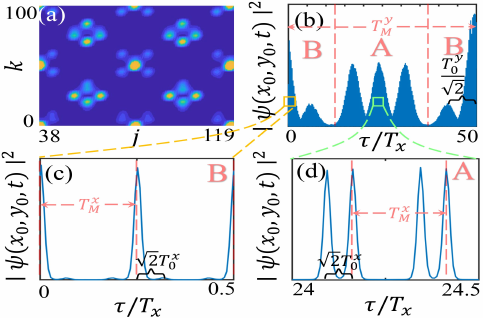}
	\caption{(a) $|\psi(x_0,y_0,t)|^2$ in $(j,k)$ temporal coordinates ($t=\Delta t_x j+\Delta t_y k+98T_y$) shows temporal Moiré superfluidity. (b) Evolution versus $\tau=t-98T_y$ reveals Moiré patterns $A/B$ at $T_y$ scale. (c,d) Higher-resolution views show nested Moiré structures at $T_x$ scale. Parameters as in Fig.~\ref{fig2}.
}
	\label{fig3}
\end{figure}

The overall structure and dynamics of temporal regional superfluid are illustrated in Fig.~\ref{fig3}, where Fig.~\ref{fig3}(b) shows the probability density $|\psi(t)|^2$ at a fixed position over $50T_x$ in continuous time, with enlarged views in Fig.~\ref{fig3}(c) and \ref{fig3}(d), while panel Fig.~\ref{fig3}(a) uses discrete time coordinates $(j,k)$ from Eqs.~(\ref{tjk})-(\ref{thetak}) to reveal a 2D moiré structure with dual patterns A and B matching the spatial distribution in Fig.~\ref{fig2}. These patterns exhibit similar variations along $j$ and $k$, but the condition $\Delta t_x \ll \Delta t_y$ places their corresponding dynamics on distinct timescales. The envelope in Fig.~\ref{fig3}(b) shows $T_y$-scale moiré dynamics with small period $T_0^y/\sqrt{2}$ and large period $T_M^y$, whereas the enlarged views in  Fig.~\ref{fig3}(c) and \ref{fig3}(d) reveal $T_x$-scale dynamics with periods $\sqrt{2}T_0^x$ and $T_M^x$ but different timing sequences for different $k$ values. 
For irrational $\Omega_x/\Omega_y$, the evolution is quasiperiodic. Critically, although $|\psi(t)|^2$ shares the quasiperiodicity of the Hamiltonian $H_{sa}(t)$, the wave function $\psi(t)$ itself exhibits quasiperiods multiplied by a factor of $|n - m|$, determined by the number of moiré cells, $2\pi/|{\mathbf a}_M^{1,2}|$. 
This folding effect arises from the interference due to phase variations between different moiré cells in the regional superfluid, which directly demonstrates discrete time translation symmetry breaking, a hallmark of time crystal behavior.

The remarkable stability of the regional superfluid time crystal against Floquet heating, a primary decoherence mechanism in periodically driven systems, can be understood within the FPS using Fermi's golden rule \cite{choudhury2015stability}. The heating rate due to atomic collisions is given by
\begin{equation}
\Gamma \approx \frac{2\pi^5 g^2 \eta^2 N^2}{\hbar L^4} \sum_f \left| \int d\bar{\theta}_{x,y}  \psi^*_f |\psi|^2 \psi \right|^2 \delta(E_f - E),
\end{equation}
where the collision interaction is approximated by the mean field potential $\sim|\psi|^2$, and the summation runs over all Floquet eigenstates of $\mathcal{H}_{sa}$. Energy conservation is evaluated approximately between the mean energy $E$ of the time crystal state $\psi$ and the eigenenergies $E_f$. Crucially, $\psi$ is a coherent superposition of several Floquet eigenstates. The destructive interference among these components makes the effective interaction potential $|\psi|^2$ significantly smaller than that for a condensate occupying a single Floquet eigenstate. This suppression sharply reduces the integral, leading to a much lower heating rate and, hence, a longer lifetime compared to conventional periodically driven condensates. A more rigorous numerical analysis based on the Floquet Fermi's golden rule in laboratory coordinates \cite{bilitewski2015scattering} is provided in Supplemental Material\cite{supplementarymaterial}, confirming this protective interference effect.

\begin{figure}[t]
	\centering
	\includegraphics[width=0.95\linewidth,height=0.6\linewidth]{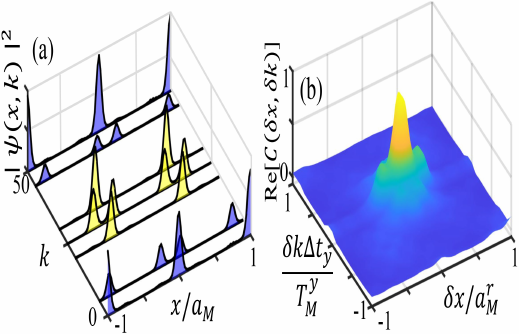}
	\caption{(a) Slices of probability density in the coordinate $(x,k)$ reveal a 2D spatiotemporal moir\'e lattice, and the two moir\'e patterns are denoted with different shading colors. (b) Spatiotemporal autocorrelation $C(\delta x,\delta k)$ in the 2D coordinate $(x,k)$ possesses a spatiotemporal coherent length up moir\'e scale. Parameters as in Fig.~\ref{fig2}. }
	\label{fig4}
\end{figure}

In addition to unveiling the 2D regional superfluid state in both spatial and temporal coordinates, our scheme also permits the emergence of a spatiotemporal regional superfluid state. This state is distinguished by a probability density distribution exhibiting a moiré pattern in the $(x,k)$ coordinates and by the presence of atomic phase coherence across both spatial and temporal moiré periods. As illustrated in Fig.~\ref{fig4}(a), for a constant time $k$, only one type of periodic density distribution is evident along the $x$ coordinate. As time $k$ varies, two distinct types of periodic density distributions alternately appear, corresponding to moir\'e patterns $A$ and $B$. An analogous alternation is observed in the temporal periodic density distribution as the position $x$ changes.
A spatiotemporal correlation $C(\delta x, \delta k)$ defined through the combination of Eqs.~(\ref{Cr}) and (\ref{Ct}) is depicted in Fig.~\ref{fig4}(b). The bulge observed in the center emphasizes that coherence can persist within a range corresponding to the spatiotemporal moiré period $a_M = \sqrt{\delta \tau^2 + \delta x^2}$, indicating a necessary trade-off between temporal delay and spatial distance when leveraging this quantum coherence.

\emph{Conclusion}---We demonstrate how ultracold atoms in a 2D deep potential well, subjected to multifrequency perturbations but requiring neither spatial periodicity nor multilayer potentials, can emulate twistronics physics. Their quantum dynamics form a 2D moiré time crystal when prepared with proper initial velocities and interatomic interaction strength, exhibiting regional superfluidity characterized by moiré-patterned probability distributions and coherence lengths across spatial, temporal, or spatiotemporal domains. 
For cold $^{39}K$ atoms in a $\sim \!100 \ \mu m$ box potential with initial velocities $v_x\sim\! 10\ m/s$ and $v_y\sim\! 0.1 \ m/s$, we obtain a $\sim\! 10 \ \mu m$ spatial moiré pattern and two distinct temporal moiré periods $T_M^x \sim\! 1 \ \mu s$ and $T_M^y \sim\! 0.1 \ ms$ \cite{PhysRevA.105.043319}, which are $|2mn/(m^2-n^2)|$ times the primary periods. Increasing the perturbation frequency multiples $n$ and $m$ while keeping them close could enhance moiré primary period separation, generating carrier-envelope structure analogous to mode-locked laser  \cite{doi:10.1126/science.288.5466.635}, and paving the way for ultrafast atom lasers \cite{ROBINS2013265}. The scheme accommodates other moiré phases through interaction and perturbation frequency tuning, extends naturally to 3D moiré crystals \cite{wang2024three}, and suggests possible spacetime crystal realizations \cite{xu2018space, vzlabys2021six, peng2022topological}.

\let\oldaddcontentsline=\addcontentsline
\renewcommand{\addcontentsline}[3]{}

\begin{acknowledgments}
\emph{Acknowledgements}---This work was supported by the Quantum Science and Technology–National Science and Technology Major Project (No. 2021ZD0303200); the National Natural Science Foundation of China (No. 12374328, No. 11974116, and No. 12234014); the Science and Technology Innovation Plan of Shanghai Science and Technology Commission (No. 24LZ1400600); the Shanghai Municipal  Science and Technology Major Project (No. 2019SHZDZX01); the National Key Research and Development Program of China (No. 2016YFA0302001); the Fundamental Research Funds for the Central Universities; the Chinese National Youth Talent Support Program, and the Shanghai Talent program.
\end{acknowledgments}

\emph{Data availability}---The data that support the findings of 
this article are not publicly available upon publication 
because it is not technically feasible and/or the cost of 
preparing, depositing, and hosting the data would be 
prohibitive within the terms of this research project. The 
data are available from the authors upon reasonable request.

\nocite{*}
\bibliography{Ref}

\renewcommand{\addcontentsline}{\oldaddcontentsline} 

\pagebreak
\clearpage
\widetext
\begin{center}
    \textbf{\large Supplemental Materials for "Atomic Regional Superfluids in Two-Dimensional Moir\'e Time Crystals"}
\end{center}

\setcounter{equation}{0}
\setcounter{figure}{0}
\setcounter{table}{0}
\setcounter{page}{1}
\setcounter{section}{1}
\makeatletter
\renewcommand{\theequation}{S\arabic{equation}}
\renewcommand{\thefigure}{S\arabic{figure}}
\renewcommand{\bibnumfmt}[1]{[#1]}
\renewcommand{\citenumfont}[1]{#1}


\tableofcontents
\section{2D Action-Angle Transform}

In the single-atom case, the atom is confined by a conservative two-dimensional potential and perturbed by a time-dependent potential. The Hamiltonian describing this system is given by
\begin{equation}
H_{sa} = \frac{p_x^2+p_y^2}{2m_{a}} + U(x,y) + V(x,y,t)
\label{Hsa}
\end{equation}
where $U(x,y)$ represents an infinite square well defined as
\begin{equation}
U(x,y) =
\begin{cases}
0 & \text{if } |x|, |y| < \frac{L}{2}, \\
\infty & \text{otherwise}\,\,\,.
\end{cases}
\end{equation}
The time-dependent perturbation is given in the form
\begin{equation}
V(x,y,t) = V_0 x^2y^2 f(t),
\end{equation}
where the function $f(t)$ describing the temporal variation is
\begin{equation}
f(t) = \cos(\omega_1 t) + \cos(\omega_2 t) + \cos(\omega_3 t) + \cos(\omega_4 t).
\end{equation}

Without the influence of perturbation, the atom's motion can be decomposed into components along the $x$ and $y$ dimensions, with the periods determined by the initial kinetic energies $E_{x}$ and $E_{y}$, respectively. This periodic motion can be transformed from the $(x, y)$ coordinate space into a 2D angular space $(\theta_x, \theta_y)$ using the action-angle transformation,
\begin{eqnarray}
I_{x} &=& \frac{L\sqrt{2 m_a E_{x}}}{\pi}, \\
\theta_{x} &=& \frac{\pi}{L} \left( x + \frac{L}{2} \right) \text{sign}(p_{x}),
\end{eqnarray}
with analogous expressions for $I_y$ and $\theta_y$ in terms of $E_y$ and $y$.
The evolution of the angles is constrained within the range $(-\pi, \pi]$ and is governed by the frequencies
\begin{equation}
    \Omega_{x,y} = \dot{\theta}_{x,y}=\frac{\partial E_{x,y}}{\partial I_{x,y}} = \frac{\pi}{L} \sqrt{\frac{2E_{x,y}}{m_a}}
\end{equation}

In angular space, the first two terms of $H_{sa}$ becomes
\begin{equation}
    H_{0}(I_x, I_y) = \frac{\pi^2(I_x^2 + I_y^2)}{2m_aL^2},
\end{equation}
and the perturbation becomes
\begin{equation}
    V(x,y,t) = \frac{V_0 L^4}{\pi^4} \left( \left| \theta_x \right| - \frac{\pi}{2}  \right)^2 \left( \left| \theta_y\right| - \frac{\pi}{2}  \right)^2 f(t) = 
    \sum_{\mathfrak{m},\mathfrak{n},l} \mathcal{V}_{\mathfrak{nm}l} e^{i(\mathfrak{n}\theta_x + \mathfrak{m}\theta_y \pm \omega_l t)},\label{Vnml}
\end{equation}
with Fourier coefficients
\begin{equation}
    \mathcal{V}_{\mathfrak{nm}l} = \frac{4V_0 L^4 f_l}{\pi^4 \mathfrak{n}^2 \mathfrak{m}^2},
\end{equation}
where $\mathfrak{n},\mathfrak{m}$ are integers, though the specific form of the potential results in only even terms appearing in the Fourier series. The full system is then described by the Hamiltonian
\begin{equation}
    H_{sa} = H_0(I_x, I_y) + V(\theta_x, \theta_y, t).
\end{equation}

\section{Secular Approximation and Floquet Phase Space}
In the absence of the periodic perturbation $V$, the action variables $I_{x,y}$ are constants of motion, and the angle variables evolve linearly in time as $\theta_{x,y} = \Omega_{x,y}t$. For a sufficiently weak perturbation $V$, the trajectories of the perturbed system remain close to the unperturbed ones. This implies that the deviations $\theta_{x,y} - \Omega_{x,y}t$ and the action variables $I_{x,y}$ become slowly varying quantities.

When the driving frequencies satisfy the resonance conditions
\begin{eqnarray}
    \omega_{1} &= n\Omega_{x} + m\Omega_{y}, \\
    \omega_{2} &= m\Omega_{x} + n\Omega_{y}, \\
    \omega_{3} &= n\Omega_{x} - m\Omega_{y}, \\
    \omega_{4} &= m\Omega_{x} - n\Omega_{y},
\end{eqnarray}
we can apply the “secular approximation”\cite{buchleitner2002non}. Under this approximation, all non-resonant terms $\mathcal{V}_{\mathfrak{nm}l} e^{i(\mathfrak{n}\theta_x + \mathfrak{m}\theta_y \pm \omega_l t)}$ oscillate rapidly and average to zero over time. Only the resonant terms, which satisfy $\mathfrak{n}\Omega_x+\mathfrak{m}\Omega_y=\omega_l$, evolve slowly and dominantly influence the long-term dynamics.
We therefore transform to a frame co-moving with the unperturbed motion by defining the slow angle variables
\begin{equation}
    \bar{\theta}_{x,y}=\theta_{x,y} - \Omega_{x,y} t.
\end{equation}
After dropping the fast-oscillating terms, the perturbation potential reduces to the effective form
\begin{equation}
    \mathcal{V}(\bar{\theta}_{x},\bar{\theta}_{y})=\frac{4V_{0}L^{4}}{\pi^{4}n^2m^2}\left[\cos(n\bar{\theta}_{x}+m\bar{\theta}_{y})+\cos(m\bar{\theta}_{x}+n\bar{\theta}_{y})+\cos(n\bar{\theta}_{x}-m\bar{\theta}_{y})+\cos(m\bar{\theta}_{x}-n\bar{\theta}_{y})\right].
\end{equation}


However, in systems with two degrees of freedom $( \theta_x, \theta_y )$, accidental resonances may arise when different mode combinations satisfy $ n\Omega_x + m\Omega_y = n^\prime\Omega_x + m^\prime\Omega_y $ for distinct integer pairs. We adopt an irrational ratio $ \Omega_x / \Omega_y $ to suppress them. Nevertheless, near-resonant terms with small $ |n\Omega_x + m\Omega_y - \omega_l| $ persist and can invalidate the secular approximation via cumulative effects. This disruption can be mitigated by a specially designed perturbation potential, such as $ V \propto x^2 y^2$. We will now employ canonical perturbation theory to identify the general design principles for such potentials.

We begin with a time-dependent canonical transformation
\begin{equation}
H_{\mathrm{sa}}(I_{x,y},\theta_{x,y},t) + \frac{\partial G}{\partial t} = \mathcal{H}_{\mathrm{sa}}(I^\prime_{x,y}, \bar{\theta}_{x,y}) + \delta\mathcal{H}_{\mathrm{sa}},
\label{cantran}
\end{equation}
where $\mathcal{H}_{\mathrm{sa}}$ is the secular Hamiltonian and $\delta\mathcal{H}_{\mathrm{sa}}$ captures residual near-resonant corrections. This framework allows us to identify parameter regimes that minimize the impact of these terms.

The generating function $G$, organized as a series in the perturbation strength $V_0$, is
\begin{equation}
G = G^{(0)} + G^{(1)},
\end{equation}
where the zeroth-order term
\begin{equation}
G^{(0)} = I_x^\prime(\theta_x - \Omega_x t) + I_y^\prime(\theta_y - \Omega_y t)
\end{equation}
transforms the system into the co-moving frame, and the first-order term 
\begin{equation}
G^{(1)} = i \sum_{\mathfrak{n,m},l}^{\prime} \left[ \frac{\mathcal{V}_{\mathfrak{nm}l}}{\mathfrak{n}\Omega_x + \mathfrak{m}\Omega_y + \omega_l} e^{i(\mathfrak{n}\theta_x + \mathfrak{m}\theta_y + \omega_l t)} + \frac{\mathcal{V}_{\mathfrak{nm}l}}{\mathfrak{n}\Omega_x + \mathfrak{m}\Omega_y - \omega_l} e^{i(\mathfrak{n}\theta_x + \mathfrak{m}\theta_y - \omega_l t)} \right]
\end{equation}
is designed to remove non-resonant perturbations\cite{lichtenberg2013regular}. The prime on the summation indicates the exclusion of resonant terms.

The generating function $G$ defines the transformation relations
\begin{eqnarray}
I_{x,y} &=& \frac{\partial G}{\partial \theta_{x,y}} = I^\prime_{x, y} + \partial_{\theta_{x,y}} G^{(1)}, \\
\bar{\theta}_{x,y} &=& \frac{\partial G}{\partial I^\prime_{x,y}} = \theta_{x,y} - \Omega_{x,y} t.
\end{eqnarray}
Substitution into Eq.~\eqref{cantran} and expansion in a power series of $\partial_{\theta_i} G^{(1)}$ gives
\begin{equation}
H_0(I_{i}^\prime)+ \partial_{I^\prime_i} H_{0}\, \partial_{\theta_i} G^{(1)}
+ \tfrac{1}{2} \partial_{I^\prime_i} \partial_{I^\prime_j} H_0 \, \partial_{\theta_i} G^{(1)}\partial_{\theta_j} G^{(1)} \cdots \,
 + V(\bar{\theta}_{i}, t) + \partial_t G^{(0)} +\partial_t G^{(1)} = \mathcal{H}_{\mathrm{sa}} + \delta\mathcal{H}_{\mathrm{sa}},
\label{cantran2}
\end{equation}
where indices $i,j \in \{x,y\}$ are summed over. The first-order terms $\partial_{I^\prime_i} H_{0}\, \partial_{\theta_i} G^{(1)}+\partial_t G^{(1)}$ are designed to cancel the non-resonant part of $V(\bar{\theta}_{x,y}, t)$. After this cancellation, the remaining time-independent part of the potential yields the secular potential $\mathcal{V}(\bar{\theta}_{x,y})$. The higher-order terms in the expansion collectively form the correction $\delta\mathcal{H}_{\mathrm{sa}}$, representing the error inherent in the secular approximation.

Owing to the weak perturbation, the new actions $I_{i}^\prime$ exhibit only small deviations from their unperturbed values $I_{i}^0$, justifying an expansion of $H_0(I_{i}^\prime)$ around $I_{i}^0$. This leads to the secular Hamiltonian
\begin{equation}
\mathcal{H}_{\mathrm{sa}} \approx \frac{P_{\bar{\theta}_x}^2 + P_{\bar{\theta}_y}^2}{2 m_\text{eff}} + \mathcal{V}(\bar{\theta}_x, \bar{\theta}_y),
\end{equation}
where we define the action deviation
\begin{equation}
P_{\bar{\theta}_{x,y}} = I_{x,y}^\prime - I_{x,y}^0
\end{equation}
as the effective momentum conjugate to the slow variables $\bar{\theta}_{x,y}$. The effective mass is given by $m_\text{eff} = \left( \partial^2 H_0 / \partial I^{\prime 2}_i \right)^{-1} = \frac{m_a L^2}{\pi^2}$. In this derivation, constant energy offsets are removed, and the linear terms in $P_{\bar{\theta}_i}$ are canceled by the contribution from $\partial_t G^{(0)}$. The phase space spanned by $(P_{\bar{\theta}_{i}}, \bar{\theta}_{i})$ is termed the Floquet phase space because quantizing these variables yields a time-independent quantum Hamiltonian, which is analogous to the quantum Floquet Hamiltonian in the high-frequency limit \cite{buchleitner2002non, sacha2016anderson}.
 
The error of the secular approximation, originating from near-resonant perturbations, is given to second order in $V_0$ by
\begin{equation}
\delta \mathcal{H}_{sa} \approx \frac{1}{2} \frac{\partial^2 H_0}{\partial I^\prime_i \partial I^\prime_j} (\partial_{\theta_i} G^{(1)} \partial_{\theta_j} G^{(1)})= \frac{\pi^2}{2m_aL^2}\sum^\prime_{\substack{\mathfrak{n,m},l \\ \mathfrak{n',m'},l'}} \frac{\mathcal{V}_{\mathfrak{n m}l}\mathcal{V}_{\mathfrak{n' m'}l'}(\mathfrak{n}\mathfrak{n'}+\mathfrak{m}\mathfrak{m'})\,e^{i[(\mathfrak{n}+\mathfrak{n'})\bar{\theta}_x+(\mathfrak{m}+\mathfrak{m'})\bar{\theta}_y+\Delta t]}}{(\mathfrak{n}\Omega_x+\mathfrak{m}\Omega_y+\omega_l)(\mathfrak{n'}\Omega_x+\mathfrak{m'}\Omega_y-\omega_{l'})} + \cdots,
\end{equation}
where the detuning is defined as $\Delta = (\omega_l - \omega_{l'}) + (\mathfrak{n}+\mathfrak{n'})\Omega_x + (\mathfrak{m}+\mathfrak{m'})\Omega_y$ (We show one representative term, others follow similarly).

A large detuning $ \Delta $ allows $ \delta\mathcal{H}_{sa} $ to be neglected due to rapid temporal oscillations. However, near-resonant terms amplify $ \delta\mathcal{H}_{sa} $ through small denominators in the summation. To ensure that $ \delta\mathcal{H}_{sa} $ remains negligible compared to the leading potential $\mathcal{V}(\bar{\theta}_x, \bar{\theta}_y) $, we derive the condition
\begin{equation}
    \frac{2L^2}{\pi^2 m_a} \frac{V_0}{\Omega_x \Omega_y} \frac{\mathcal{F}_{\mathfrak{n m}l} \mathcal{F}_{\mathfrak{n' m'}l'}}{\mathcal{F}_{n m l}} (\mathfrak{n}\mathfrak{n'} + \mathfrak{m}\mathfrak{m'}) \ll \delta \delta',
\end{equation}
where $ \delta = |\mathfrak{n} + \mathfrak{m}(\Omega_y/\Omega_x) \pm \omega_l/\Omega_x| $ and $ \delta' = |\mathfrak{n'}(\Omega_x/\Omega_y) + \mathfrak{m'} \pm \omega_{l'}/\Omega_y| $ are the normalized detunings, and $ \mathcal{F}_{\mathfrak{n m} l} = \mathcal{V}_{\mathfrak{n m} l} / (V_0 L^4) $ denotes the normalized Fourier coefficients of the perturbation potential (with $ \mathcal{F}_{n m l} $ corresponding to the resonant terms satisfying $ n\Omega_x + m\Omega_y = \omega_l $).

To suppress the left-hand side of this inequality, one can employ higher initial kinetic energy and a smaller perturbation strength for the cold atoms to keep the prefactor $ 2V_0 L^2 / (\pi^2 m_a \Omega_x \Omega_y) $ small. Alternatively, one can design perturbation potentials $ V(x,y,t) $ with specific spatial profiles that yield rapidly decaying Fourier coefficients $ \mathcal{F}_{\mathfrak{n m} l} $ as $ |\mathfrak{n}|, |\mathfrak{m}| $ increase. For example, the quartic spatial form $ V \propto x^2 y^2 $ used here leads to $ \mathcal{F}_{\mathfrak{n m} l} \sim \mathfrak{n}^{-2} \mathfrak{m}^{-2} $. A potential with a higher spatial power would be more effective but requires a more sophisticated experimental design.

\section{Heating Analysis with the Floquet Fermi Golden Rule}

Floquet heating, the primary decoherence mechanism in periodically driven quantum many-body systems, arises largely from two-body collisions that induce transitions between Floquet states. Time crystals are notable for their resistance to this heating. Here, we estimate the heating rate using Fermi's golden rule within the quantum Floquet framework, complementing the semiclassical Floquet phase space approach in the main text. We show that the heating resistance of our Moiré time crystal, which is a superposition of Floquet states, is significantly stronger than in conventional cold atom systems that typically occupy a single Floquet state.

For a Hamiltonian  $H(t) = H_0 + V(t)$ driven at multiple incommensurate frequencies, the time-quasiperiodic dynamics admit solutions of the generalized Floquet form \cite{peng2018time,martin2017topological, casati1989anderson, ho1983semiclassical}:
\begin{equation}
\Psi_\alpha(\mathbf{r},t) = e^{-i\epsilon_\alpha t/\hbar} \Phi_\alpha(\mathbf{r}, t) = e^{-i\epsilon_\alpha t/\hbar} \sum_{\bm{l},n} c_{\bm{l},n}^\alpha e^{-i\bm{l} \cdot \bm{\omega} t} \varphi_n(\mathbf{r}),
\end{equation}
where $\epsilon_\alpha$ is the quasienergy and $\alpha$ labels the Floquet states. The Floquet mode $\Phi_\alpha(\mathbf{r}, t)$ shares the quasiperiodicity of $V(t)$ and is expanded in a multi-frequency Fourier series, with its spatial dependence expressed via the eigenfunctions of $H_0$. Here, $\bm{\omega} = (\omega_1, \omega_2, \ldots)$ is the frequency vector, $\bm{l} = (l_1, l_2, \ldots) $ is an integer vector indexing the Fourier harmonics, and $n$ labels the eigenstates of $H_0$.

Departing from the conventional Floquet Fermi's golden rule, we consider atoms are initially in a time crystal state, specifically a superposition of Floquet states $\Psi_i = \sum_\alpha b_\alpha \Psi_\alpha$, where the coefficients $b_\alpha$ are determined by finding the ground state of the Floquet Hamiltonian that includes the collision interaction $H_{\mathrm{int}}$. Collisions can then induce transitions to a final Floquet state $\Psi_f$ through the absorption of energy quanta $\hbar \bm{m} \cdot \bm{\omega} = \hbar(m_1\omega_1 + m_2\omega_2 + \cdots)$.
Following the approach in Ref. \cite{bilitewski2015scattering,ikeda2021fermi}, the transition amplitude is
\begin{equation}
A_{fi}^{\bm{\omega}}(t) = \sum_{\alpha,\bm{m}} b_{\alpha} I_{f,\alpha,\bm{m}} \frac{e^{-i(\epsilon_\alpha - \epsilon_f - \bm{m} \bm{\omega} \hbar)t/\hbar} - 1}{\epsilon_\alpha - \epsilon_f - \bm{m} \bm{\omega} \hbar},
\end{equation}
with the transition matrix element
\begin{equation}
I_{f,\alpha,\bm{m}} = \sum_{n,n^\prime,\bm{l}} c_{n^\prime,\bm{l}+\bm{m}}^{f*} c^\alpha_{n,\bm{l}} \int d\mathbf{r} \varphi_{n^{\prime}}^* H_{\mathrm{int}} \varphi_n.
\end{equation}
The Floquet heating rate, given by the sum of steady-state transition probabilities to all possible final Floquet states, is approximated as
\begin{equation}
\Gamma = \sum_f \lim_{t\rightarrow\infty}\frac{|A^{\bm{\omega}}_{fi}(t)|^2}{t} \approx \frac{2\pi}{\hbar}\sum_{f, m,\omega_k} \delta(\bar{\epsilon}_i - \epsilon_f - m \hbar \omega_k) \left| \sum_{\alpha} b_{\alpha} I_{f,\alpha,m} \right|^2.
\end{equation}
Two approximations lead to this form. First, in the regime of weak multi-frequency driving, the system responds linearly \cite{zhao2021random}, meaning energy is absorbed from each driving frequency $\omega_k$ independently. This permits a simplification of the summation over the vector $\bm{m} $ into separate sums over the scalar $m$ and $\omega_k$. Second, the small quasienergy fluctuation among the Floquet state components of the initial state $\Psi_i$, satisfying $\delta\epsilon_i\ll\hbar\omega_k$, justifies replacing $\epsilon_\alpha$ with its mean value $\bar{\epsilon}_i$ outside the summation over $\alpha$. Notably, for a single Floquet initial state (where the $\alpha$-summation vanishes), our general expression simplifies to the standard Floquet heating rate given in \cite{bilitewski2015scattering, ikeda2021fermi}.

The discrete sum over final states $ \sum_f $ is converted to an integral, $ \sum_f \rightarrow \int \frac{df}{d\epsilon_f} d\epsilon_f $. In the weak driving limit ($ \mathcal{V} \ll \hbar\Omega_{x,y} $), the density of states $ \frac{df}{d\epsilon_f} $ can be approximated by that of an infinite square well, $ \frac{m_a L^2}{2\pi\hbar^2} $. Furthermore, leveraging the coherence of the condensate, we model atomic collisions via a mean-field interaction potential $ H_{\text{int}} \approx g_{2D}N |\Psi_i|^2 $. The heating rate $ \Gamma $ is then computed numerically.

\begin{figure}[h]
	\centering
	\includegraphics[width=0.5\linewidth, height=0.3\linewidth]{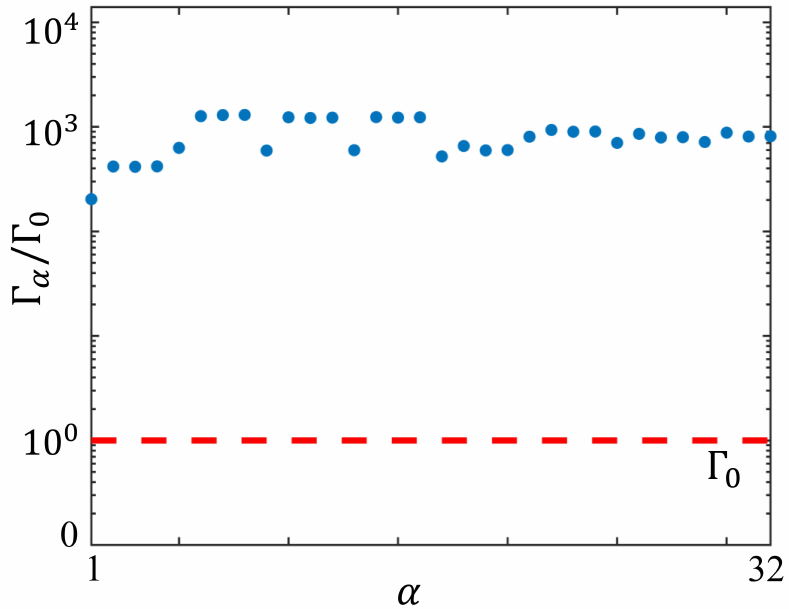}
    \caption{ Heating rates $\Gamma_\alpha$ (blue dots) for the lowest $32$ single Floquet states, whose superposition forms the Moiré time crystal state with heating rate $\Gamma_0$ (dashed line). Parameters are the same as in the main text.
    }
\end{figure}

As shown in Fig. S1, the heating rates for different single Floquet states $ \Psi_\alpha $ vary, yet they are consistently about three orders of magnitude larger than that for the Moiré time crystal state, a superposition of Floquet states. Specifically, for a condensate in a single Floquet state, the lifetime $ \tau \sim N\Gamma^{-1} $ is limited to about $ 10^2 N T_y\sim N\times10^{-2}s $. In contrast, the Moiré time crystal exhibits a drastically prolonged lifetime of $ \sim 10^5 N T_y \sim N\times 10s$ for the parameters in the main text. This profound enhancement stems from destructive quantum interference among the constituent Floquet states, which suppresses the key scattering channels and strongly reduces the heating rate. This directly manifests the time crystal's non-trivial quantum nature and its inherent protection against decoherence and heating.

\section{Analysis of the Phase Correlation Length within Bogoliubov Theory}

In this section, we employ Bogoliubov theory to provide a qualitative analysis of the attenuation behavior observed in the autocorrelation of the Moiré time crystal, as obtained from numerical simulations in Fig. 2 of the main text.

To reconcile the spatial and temporal dimensions, we begin our analysis within the framework of Floquet phase space. The first-order autocorrelation function is defined as
\begin{equation}
C^{(1)}(\delta\bm{\bar\theta}) = \int \langle \hat{\psi}^\dagger(\bm{\bar\theta}) \hat{\psi}(\bm{\bar\theta} + \delta\bm{\bar\theta}) \rangle d\bm{\bar{\theta}},
\end{equation}
where $\bm{\bar\theta} = (\bar{\theta}_x, \bar{\theta}_y)$ is a vector in the two-dimensional phase space, and the integration is carried out over both dimensions.

To analyze correlations specific to the lattice geometry, we express the autocorrelation function in terms of quasi-momentum states by expanding the field operator in the Bloch basis
\begin{equation}
\hat{\psi}(\bm{\bar\theta}) = \sum_{\bm{k}} u_{\bm{k}}(\bm{\bar\theta}) e^{i\bm{k} \cdot \bm{\bar\theta}} \hat{c}_{\bm{k}},
\end{equation}
where $\bm{k} = (k_x, k_y)$ is the quasi-momentum. Substituting this expansion yields
\begin{equation}
C^{(1)}(\delta\bm{\bar\theta}) = \sum_{\bm{k}} I_{\bm{k}}(\delta\bm{\bar\theta}) e^{i \bm{k} \cdot \delta\bm{\bar\theta}} \langle \hat{c}^\dagger_{\bm k} \hat c_{\bm k} \rangle.
\label{C1k}
\end{equation}
This representation generalizes the standard Fourier relation through a modulation factor 
\begin{equation}
    I_{\bm{k}}(\delta\bm{\bar\theta}) = \int u_{\bm k}^*(\bm{\bar\theta})u_{\bm k}(\bm{\bar\theta}+\delta\bm{\bar\theta})d\bm{\bar\theta},
\end{equation}
which quantifies the degree of atomic localization within a unit cell.

For a shallow lattice and for states with low quasi-momentum $\bm k$, which dominate the long-range behavior, the factor $I_{\bm{k}}(\delta\bm{\bar\theta})$ varies weakly with $\bm{k}$ and can therefore be taken outside the summation. In this limit, the autocorrelation function simplifies to a Fourier transform of the quasi-momentum density distribution, modulated by an envelope function that depends on the periodic atomic density distribution.

To analyze the decay characteristics of the autocorrelation function, we focus on the non-condensed portion of the atoms by approximating $\langle \hat{c}_0 \rangle \approx \sqrt{N_0}$ and treating operators $\hat{c}_{\bm{k} \neq 0}$ as small fluctuations. The behavior of these fluctuations can be described approximately by the quadratic Hamiltonian
\begin{equation}
\delta \hat{H} = \sum_{\bm{k} > 0} \left[ (\varepsilon_{\bm{k}}^0 + U_0 n_0) \left( \hat{c}^\dagger_{\bm{k}} \hat{c}_{\bm{k}} + \hat{c}^\dagger_{-\bm{k}} \hat{c}_{-\bm{k}} \right) + U_0 n_0 \left( \hat{c}^\dagger_{\bm{k}} \hat{c}^\dagger_{-\bm{k}} + \hat{c}_{\bm{k}} \hat{c}_{-\bm{k}} \right) \right],
\end{equation}
where $U_0 = \frac{\pi^4 }{L^4}g_{2D}\int  d\bm{\bar\theta} |w(\bm{\bar\theta})|^4$ is the on-site interaction strength, with $w(\bm{\bar\theta})$ being the Wannier function of the primary lattice. The condensate density is given by $n_0 = N_0 / (2\pi/a_0)^2$, and $\varepsilon_{\bm{k}}^0$ represents the single-particle energy.

The quadratic Hamiltonian is diagonalized via the Bogoliubov–de Gennes transformation, $\hat{c}_{\bm k}=\mu_{\bm k}^*\hat{b}_{\bm k}-\nu_{\bm k}\hat{b}_{-\bm k}^\dagger$, yielding
\begin{equation}
    \delta\hat{H}=\sum_{\bm k\neq0}\varepsilon_{\bm k}(\hat{b}_{\bm k}^\dagger \hat{b}_{\bm k}+\frac{1}{2}),
\end{equation}
where the quasiparticle energy and transformation coefficients are given by
\begin{eqnarray}
\varepsilon_{\bm k}&=&\sqrt{\varepsilon_{\bm k}^0(\varepsilon_{\bm k}^0+2U_0 n_0)}, \label{ek}\\
|\nu_{\bm k}|^2&=&|\mu_{\bm k}|^2-1=\frac{\varepsilon_{\bm k}^0+U_0 n_0}{2\varepsilon_{\bm k}}-\frac{1}{2}.\label{nuk}
\end{eqnarray}
At zero temperature, the system contains no excitations, and the ground state corresponds to the Bogoliubov vacuum, satisfying $\langle \hat{b}_{\bm{k}}^\dagger \hat{b}_{\bm{k}} \rangle = 0$. Using the Bogoliubov transformation, we obtain $\langle \hat{c}_{\bm{k}}^\dagger \hat{c}_{\bm{k}} \rangle = |\nu_{\bm{k}}|^2$. 
Substituting this result along with Eqs. (\ref{C1k}), (\ref{ek}), and (\ref{nuk}) yields the approximate form of the correlation function:
\begin{equation}
C^{(1)}(\delta\bm{\bar\theta}) \approx I(\delta\bm{\bar\theta})\int^{\bm k_{B}}_{0} d\bm k e^{i \bm{k} \cdot \delta\bm{\bar\theta}} \left( \frac{\varepsilon_{\bm{k}}^0 + U_0 n_0}{2\sqrt{\varepsilon_{\bm{k}}^0 (\varepsilon_{\bm{k}}^0 + 2U_0 n_0)}} - \frac{1}{2} \right),
\label{C1B}
\end{equation}
where the summation over $\bm{k}$ has been approximated by an integral restricted to the first Brillouin zone of the primary lattice.

The long-range behavior of $ C^{(1)}(\delta\bm{\bar\theta}) $ is governed by small-$\bm{k}$ components, a direct consequence of its Fourier structure. Although the primary lattice is shallow enough to sustain atomic condensation, the superimposed moiré lattice folds the primary Brillouin zone, opening narrow band gaps at small quasi-momenta. These low-energy spectral discontinuities in $\varepsilon_{\bm{k}}^0$ thus lead to a significantly faster decay of correlations at large $\delta\bm{\bar\theta}$ compared to the gapless case.

To capture the dominant effect, we focus on the lowest Moiré band gap $\Delta_M$,  which occurs at the boundary of the first Moiré Brillouin zone, located at $\bm{k}_M=\bm{k}_B/M$, where $M=a_M/a_0$ is the ratio of the Moiré period to the primary lattice period. The flat dispersion for $k < k_M$ allows us to approximate the spectrum for $k > k_M$ by a quadratic expansion,
\begin{equation}
\varepsilon_{\bm{k}}^0 \approx \Delta_M + \frac{\hbar^2}{2m^*} |\bm{k} - \bm{k}_M|^2,
\end{equation}
where the effective mass $m^*$ is related to the primary lattice's depth and period.

We now substitute this into Eq. (\ref{C1B}) and change variables to $\bm{q} = \bm{k} - \bm{k}_M$, neglecting the spectral anisotropy for simplicity. This leads to an effectively one-dimensional integral for the correlation function in the long-wave limit,
\begin{equation}
C^{(1)}(\delta\bm{\bar\theta}) \approx A\,I(\delta\bm{\bar\theta}) e^{i \bm{k}_M \cdot \delta\bm{\bar\theta}} \int_0^{k_B - k_M} \frac{ \mathcal{J}_0 (q |\delta\bm{\bar\theta}|)}{\sqrt{k_\Delta^2 + q^2}} q dq,
\end{equation}
where $\mathcal{J}_0(x)$ is the zeroth-order Bessel function resulting from the angular integration in polar coordinates. The coefficient $A$ and the momentum scale $k_\Delta$ are given by
\begin{eqnarray}
    A &=& \frac{\Delta_M+U_0n_0}{2\sqrt{\frac{\hbar^2}{m^*}(\Delta_M+U_0n_0)}}, \\
k_\Delta &=& \sqrt{\frac{\Delta_M(\Delta_M+2U_0n_0)m^*}{(\Delta_M+U_0n_0)\hbar^2}}.
\end{eqnarray}

Owing to the rapid decay of the Bessel function $\mathcal{J}_0(q|\delta\bm{\bar\theta}|)$ for large $q$, the upper limit of the integral can be safely extended to infinity. This allows us to obtain an analytical expression for the autocorrelation function
\begin{equation}
C^{(1)}(\delta\bm{\bar\theta}) \approx A \,I(\delta\bm{\bar\theta}) e^{i \bm{k}_M \cdot \delta\bm{\bar\theta}} \, \frac{e^{-k_\Delta |\delta\bm{\bar\theta}|}}{|\delta\bm{\bar\theta}|},
\end{equation}
which exhibits a composite decay modulated by periodic oscillations as $|\delta\bm{\bar\theta}|$ increases.

The coherence length can thus be estimated as $\xi \sim k_\Delta^{-1}$. For separations much smaller than this length, $|\delta\bm{\bar\theta}| \ll k_\Delta^{-1}$, the slower power-law decay $\sim |\delta\bm{\bar\theta}|^{-1}$ dominates, permitting the persistence of phase coherence and supporting the existence of a condensate. In contrast, for $|\delta\bm{\bar\theta}| \gg k_\Delta^{-1}$, the faster exponential decay prevails, signifying the loss of long-range order and the absence of a condensate. Using the mapping between phase space and the laboratory frame, $\bm{\bar\theta}(x, y, t)$, the corresponding spatial and temporal coherence lengths can be determined.

\begin{figure}[h]
	\centering
	\includegraphics[width=0.6\linewidth, height=0.35\linewidth]{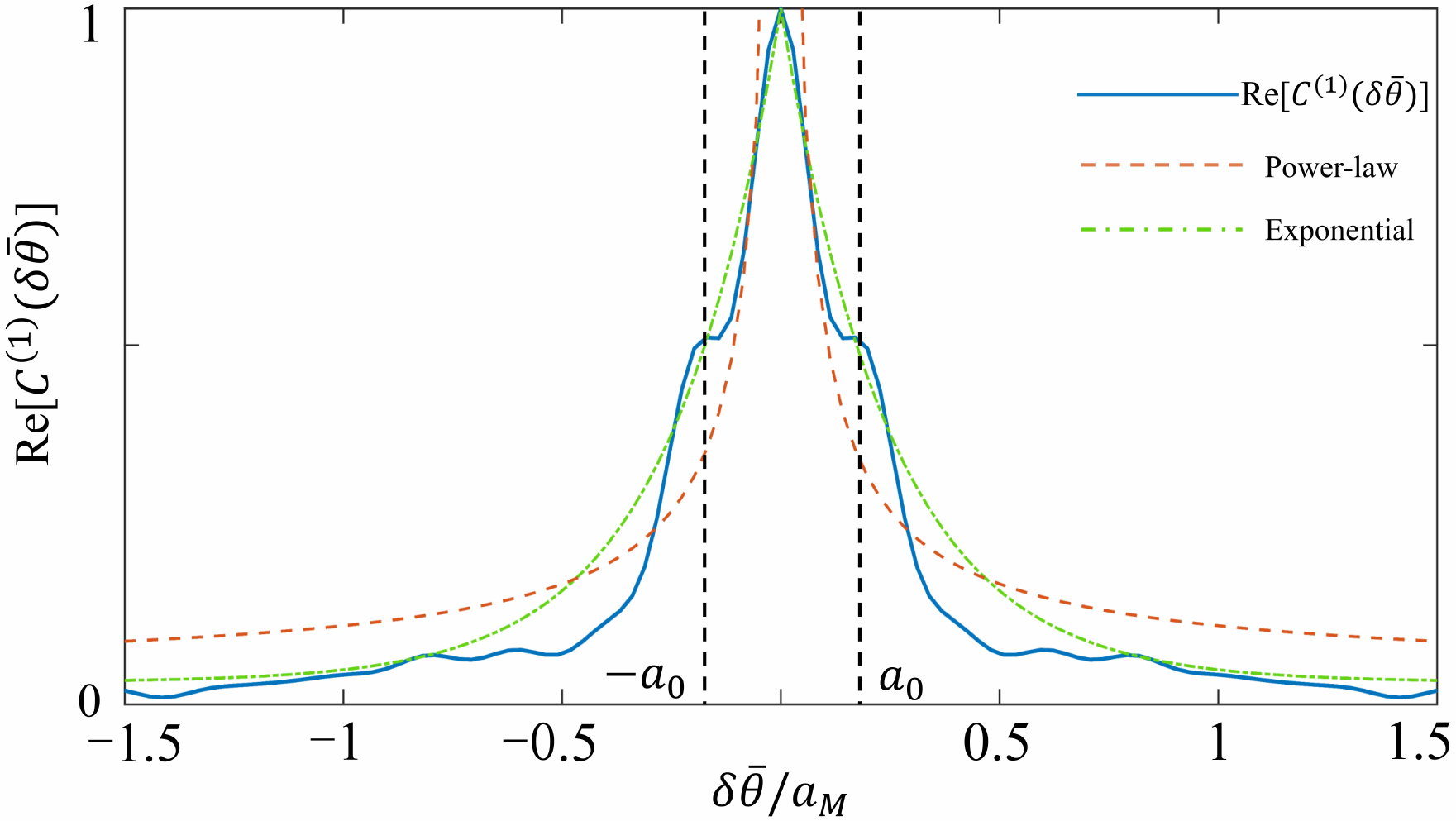}
    \caption{Autocorrelation function in FPS, showing a crossover from power-law to exponential decay at half the Moiré period.}
\end{figure}

By applying the approximated expression for the effective mass $ m^* = \frac{\hbar^2}{2 J a_0^2} $, the estimation of the tunneling rate between primary cells $ \frac{J}{E_r} \approx \frac{4}{\sqrt{\pi}} \left( \frac{\mathcal{V}}{E_r} \right)^{3/4} e^{-2\sqrt{\frac{\mathcal{V}}{E_r}}} $, and the Moiré band gap $ \Delta_M \approx \frac{2 \mathcal{V}}{M} $, where $ \frac{\mathcal{V}}{E_r} $ denotes the normalized lattice depth with respect to the recoil energy, we can roughly estimate the coherence length. For the parameters used in our numerical simulations (Fig. 2(a, b) in the main text), namely $ \frac{\mathcal{V}}{E_r} = 0.35 $, atomic interaction strength $ U_0 n_0 = 0.03 E_r $, and Moiré ratio $ M = \frac{a_M}{a_0} = 5 $, the coherence length $ k_\Delta^{-1} \approx 0.5 a_M $ is found to be larger than the primary lattice period but smaller than the Moiré period, consistent with the characteristic feature of regional superfluids.

We numerically calculate the autocorrelation function in the FPS, as shown in Fig. S2. For separations $|\delta \bar{\theta} | < 0.5a_M$, the correlation decays in a manner close to a power law, aside from side peaks at $\pm a_0$ induced by density modulation. Beyond $0.5a_M$, the decay deviates from the power-law behavior and approaches an exponential form, consistent with our theoretical analysis.

\section{Dimensional Reduction and Healing of the Condensate}

The system is reduced from 3D to 2D by applying a strong confining potential along the $z$-direction. When the energy scales associated with atomic motion and periodic driving in the $xy$-plane are much smaller than the $z$-direction excitation energy, $\hbar\omega_z$ (where $\omega_z$ is the trapping frequency along $z$), a quasi-2D atomic condensate is realized. Its mean-field dynamics are then accurately described by a 2D Gross-Pitaevskii equation
\begin{equation}
i\hbar \frac{\partial \psi}{\partial t} = \left[ -\frac{\hbar^2}{2m_a} \nabla_{\text{2D}}^2 + U_{\text{2D}}(x, y) + V_{\text{2D}}(x, y,t) + g_{\text{2D}}N |\psi|^2 \right] \psi.
\end{equation}
The effective 2D interaction coefficient,  $g_{\text{2D}}$, remains constant provided the system stays in the motional ground state along the  $z$-direction. It is related to the 3D coefficient $g_{\text{3D}} = 4\pi\hbar^2 a_s / m_a$ by $g_{\text{2D}} = \eta\, g_{\text{3D}}$, where the reduction factor $\eta = \int |\psi_{z0}(z)|^4 dz$ is computed from the ground-state wavefunction $\psi_{z0}(z)$ of the $z$-confinement. In the regime where the oscillator length $a_z = \sqrt{\hbar / (m_a \omega_z)}$ is much larger than the $s$-wave scattering length $a_s$ (typically $10^{-9}m\sim10^{-10}m$), this gives $g_{\text{2D}} = 2\sqrt{2\pi} \hbar^2 a_s / (m_a a_z)$ and $\eta^{-1}=\sqrt{2\pi}a_z$. For the case $a_z \sim a_s$, corrections beyond this approximation, which account for energy-dependent scattering, can be found in Ref. \cite{lee2002energy}.

Unlike a stationary condensate in the ground state of the 2D trap $U_{\text{2D}}(x, y)$, the atomic time crystal is prepared using higher excited states to prolong coherence under the periodic drive $V_{\text{2D}}(x, y, t)$. This requires stronger axial confinement ($\omega_z \gg \Omega_{x,y}$) than in conventional 2D condensates. The realization of effective 2D temporal degrees of freedom further necessitates a large frequency separation between in-plane modes, e.g., $\Omega_x \sim 1\ \text{kHz}$ and $\Omega_y \sim 10\ \text{Hz}$. In this case, a trapping frequency $\omega_z \sim 1\ \text{MHz}$ is sufficient. However, achieving even longer coherence times with higher excitations, such as $\Omega_x \sim 100\ \text{kHz}$ and $\Omega_y \sim 1\ \text{kHz}$, would require $\omega_z \sim 1000\ \text{MHz}$, corresponding to a confinement length $a_z \sim 10^{-9}\ \text{m}$. This poses an experimental challenge and violates the condition $a_z \gg a_s$ for energy-independent scattering.

A further experimental constraint arises from the healing length $\xi$, the scale for superfluid density recovery from the boundary. Its estimate follows from equating the Floquet kinetic energy of order $\hbar^2/(2m_{\text{eff}}\xi^2)$ to the interaction energy $\tilde{n}g_{\text{2D}}$ with $\tilde{n}\approx N/L^2$, leading to
\begin{equation}
\xi = \frac{\hbar}{\sqrt{2m_{\text{eff}} \tilde{n} g_{\text{2D}}}} = \sqrt{\frac{a_z\pi}{4\sqrt{2} a_s \tilde{n} L^2}}. 
\end{equation}
Since the regional superfluid is confined within a single Moiré supercell, the healing length must satisfy the condition $\xi \ll a_M$. 
For the parameters used in Fig. 2 of the main text, we find $\xi\approx 0.6a_0$, which is considerably smaller than both the Moiré supercell period $a_M$ and the phase correlation length $k_\Delta^{-1}$, thereby ensuring the superfluid is robust against localized perturbations.

\section{Other Realizable Spacetime Moiré Lattice Geometries}

\begin{figure}[h]
	\centering
	\includegraphics[width=0.6\linewidth, height=0.35\linewidth]{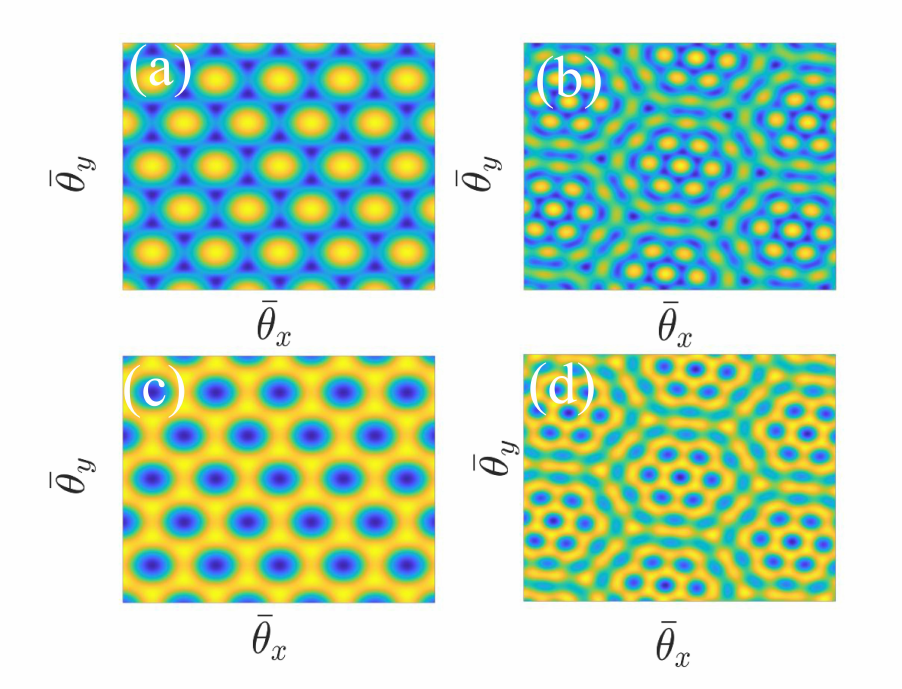}
    \caption{ (a) Honeycomb and (c) triangular lattices in Floquet phase space. (b), (d) The corresponding Moiré superlattices generated with a twist angle of $\tan\alpha = 9/41$. In all panels, the color scale represents the potential strength, ranging from low (blue) to high (yellow). The parameters used are $(n, m, n', m', s) = (12, 7, 40, 9, 41)$.}
\end{figure}

Beyond the square lattice and its Moiré superlattices discussed in the main text, a variety of other lattice geometries can be realized in the Floquet phase space by engineering the combination of perturbation frequencies $\omega_l$ in the potential $V(\theta_x,\theta_y,t)$. This potential is expressed in action-angle coordinates as
\begin{equation}
V(\theta_x,\theta_y,t) = \sum_{\mathfrak{m},\mathfrak{n},l} \mathcal{V}_{\mathfrak{nm}l} e^{i(\mathfrak{n}\theta_x + \mathfrak{m}\theta_y \pm \omega_l t)}.
\end{equation}
A honeycomb lattice, for instance, can be constructed by applying three frequencies that respectively satisfy the resonance conditions
\begin{eqnarray}
\omega_{1} &=& n\Omega_{x} + m\Omega_{y}, \\
\omega_{2} &=& n\Omega_{x} - m\Omega_{y}, \\
\omega_{3} &=& 2m\Omega_{y}.
\end{eqnarray}
The ratio $n/m=12/7$ is chosen to approximate the irrational number $\sqrt{3}$ characteristic of a standard honeycomb lattice. Applying the secular approximation to select the resonant terms in $\mathcal{V}_{\mathfrak{nm}l}$ projects the effective potential in the Floquet phase space as a sum of three cosine terms
\begin{equation}
    \mathcal{V}_{\text{honey}}(\bar{\theta}_{x},\bar{\theta}_{y})=V_0\left[\cos(n\bar{\theta}_{x}+m\bar{\theta}_{y})+\cos(n\bar{\theta}_{x}-m\bar{\theta}_{y})+\cos(2m\bar{\theta}_{y})\right],
\end{equation}
where the lattice depth $V_0$ is set by the corresponding Fourier coefficients of these resonant terms.

Constructing the same lattice geometries with a twist is straightforward. The coordinates are transformed by a rotation angle $\alpha$ as follows
\begin{eqnarray}
    \bar{\theta}_{x}^\prime &= \cos\alpha \bar{\theta}_{x} - \sin\alpha\bar{\theta}_{y}, \\
    \bar{\theta}_{y}^\prime &= \sin\alpha \bar{\theta}_{x} + \cos\alpha \bar{\theta}_{y}, 
\end{eqnarray}
where the twisted angle $\alpha$ is defined by the integers $m^\prime$ and $n^\prime$ through the relation $\tan\alpha=m^\prime/n^\prime$.

A Moiré honeycomb superlattice can then be constructed by superposing two honeycomb lattices twisted by this angle $\alpha$. This is achieved by applying six perturbation frequencies that respectively satisfy the following resonance conditions
\begin{eqnarray}
    \omega_{1} &=& s(n\Omega_{x} + m\Omega_{y}), \\
    \omega_{2} &=& s(n\Omega_{x} - m\Omega_{y}), \\
    \omega_{3} &=& 2sm\Omega_{y}, \\
    \omega_{4} &=& (n\cos\alpha+m\sin\alpha)s\Omega_{x} + (m\cos\alpha- n \sin\alpha)s\Omega_{y}, \\
    \omega_{5} &=& (n\cos\alpha- m \sin\alpha)s\Omega_{x} -(m \cos\alpha+ n \sin\alpha)s\Omega_{y}, \\
    \omega_{6}&=& 2sm(\sin\alpha\Omega_{x}+\cos\alpha\Omega_y).
\end{eqnarray}
Here, the scale factor $s=\sqrt{{m^\prime}^2+{n^\prime}^2}$ is introduced to ensure that the coefficients of $\Omega_{x,y}$ are integers. 
The resulting effective potential in the Floquet phase space is given by
\begin{eqnarray}   \mathcal{V}_{\text{honey}}^\prime(\bar{\theta}_{x},\bar{\theta}_{y})&=&V_0\left[\cos(sn\bar{\theta}_{x}+sm\bar{\theta}_{y})+\cos(sn\bar{\theta}_{x}-sm\bar{\theta}_{y})+\cos(2sm\bar{\theta}_{y})\right]+V_0\cos(2mm^\prime\bar{\theta}_{x}+2mn^\prime\bar{\theta}_{y}) \nonumber\\
     &+&V_0\cos\left[(nn^\prime+m^\prime m)\bar{\theta}_{x}+(mn^\prime-m^\prime  n)\bar{\theta}_{y})\right]+
    V_0\cos\left[(nn^\prime-m^\prime m)\bar{\theta}_{x}-(mn^\prime+m^\prime n)\bar{\theta}_{y})\right].
\end{eqnarray}

An initial phase of $\pi$ in the driving function can flip the sign of the effective potential, converting a honeycomb lattice into a triangular one. The potentials are given by $\mathcal{V}_{\text{triang}} = -\mathcal{V}_{\text{honey}}$. Likewise, the Moiré honeycomb lattice becomes a Moiré triangular lattice, $\mathcal{V}_{\text{triang}}^\prime = -\mathcal{V}_{\text{honey}}^\prime$.
Figure S3 displays these lattices for a twist angle $\tan\alpha = 9/41$. (a,b) the honeycomb and Moiré honeycomb lattices, and (c,d) their triangular counterparts.

We note that the permissible values of the twist angle $\alpha$ are limited by the constraint that the Fourier orders must be integers. In addition, possible accidental and near-resonances, as analyzed in the former section, must also be considered.


%

\end{document}